\documentclass[10pt, conference]{IEEEtran}
\IEEEoverridecommandlockouts
\usepackage{cite}
\usepackage{amsmath,amssymb,amsfonts}
\usepackage{algorithmic}
\usepackage{graphicx}
\usepackage{textcomp}
\usepackage{xcolor}
\def\BibTeX{{\rm B\kern-.05em{\sc i\kern-.025em b}\kern-.08em
    T\kern-.1667em\lower.7ex\hbox{E}\kern-.125emX}}

\usepackage{subcaption}
\usepackage{subfiles}
\usepackage[inline]{enumitem}

\usepackage{arydshln} 
\usepackage{url}
\usepackage{epstopdf}
\usepackage{tikz}
\usetikzlibrary{automata,arrows,positioning,calc,%
shapes,chains}
\usepackage{mathtools}
\usepackage{texdef2015}
\allowdisplaybreaks

\renewcommand{\vec}[1]{\begin{bmatrix} #1\end{bmatrix}}

\newcommand{\age}{\Delta}

\newlength{\swwidth}

\usepackage[T1]{fontenc}

\begin{document}

\title{Timely Mobile Routing: An Experimental Study
}

\author{\IEEEauthorblockN{Vishakha Ramani,
Jiachen Chen,
Roy D. Yates}
\IEEEauthorblockA{WINLAB, Rutgers University \\
Email: \{vishakha, jiachen, ryates\}@winlab.rutgers.edu
}}

\maketitle

\begin{abstract}
Time-critical  cyber-physical applications demand the timely delivery of information.
In this work, we employ a high-speed packet processing testbed to quantitatively analyze a packet forwarding application running on a shared memory multi-processor architecture, where efficient synchronization of concurrent access to a Forwarding Information Base is essential for low-latency and timely delivery of information. While modern packet processing frameworks are optimized for maximum packet throughput, their ability to support timely delivery remains an open question. Here we focus on the age of information performance issues induced by throughput-focused packet processing frameworks. Our results
underscore the importance of careful selection of offered load parameters and concurrency constructs in such frameworks. 
\end{abstract}


\section{Introduction}



In various cyber-physical systems and applications, sources generate {\em status updates}, time-stamped measurements of a random process of interest,  that  are sent to one or more monitors through a network. 
These applications often require timely updates at the monitors; some applications need  the age of a monitor's most recent received update to be on the order of 1-10ms. For example, 
in robotic telesurgery, a surgeon needs timely feedback on the robotic arm since stale feedback can lead to organ injuries and other ramifications \cite{dtremotesurgery},\cite{surgendoscopy}. In the Vehicle-to-Infrastructure scenario, vehicles require pre-crash sensor warning to be delivered within $20$ms.

This necessitates low latency transmission of each update.
Furthermore, because delays accumulate across multihop network paths, delays of 10-100 $\mu$s within each network node cannot be neglected. It therefore becomes imperative to 
scrutinize current network architectures and packet processing frameworks in terms of their feasibility to satisfy  stringent real-time and freshness requirements. 

We note that there is a significant literature on low-latency packet processing frameworks \cite{belay2014ix, pontarelli-flowblaze}. However, there has been a lack of studies on the ability of these frameworks to support timely updates. There also have been many analytical studies addressing the timeliness of status updates in idealized models of queues, networks and other service facilities;
see the surveys  \cite{kosta2017age,Yates-SBKMU-2021jsac-survey} and references therein. However, there has been limited  examination of update timeliness in practical systems \cite{Kadota2020age,Kadota2021wifresh,Shreedar-KY-wowmom2019}.  

In this work, we design and implement a packet forwarding experiment on a high-speed testbed employing the Data Plane Development Kit (DPDK)\footnote{DPDK is an open-source software project managed by the Linux Foundation and is widely used 
in 
data centers and core routers. 
} \cite{dpdk} packet processing framework and use
Age of Information (AoI) metric \cite{Kaul-YG-infocom2012} to evaluate two key issues that influence timeliness:
\begin{enumerate*}
    \item batch packet admission procedures of DPDK that cause input queueing, and
    \item synchronization primitives that regulate concurrent access to the Forwarding Information Base (FIB). 
\end{enumerate*}
We now present these two challenges in detail.

\subsection{Impact of input queueing}
From the outset of AoI analysis of updating systems, the value of ``bufferless'' mechanisms that discard old updates and/or give priority to 
fresher updates has been recognized \cite{Kaul-YG-infocom2012,Costa-CE-IT2016management}. In the practical context of DPDK, this would correspond to DPDK input rings that hold just a single packet. However, because DPDK  aims to maximize packet throughout, it uses large rings to absorb traffic bursts and does not support small buffer configurations. 
While this helps to avoid packet dropping (and consequent TCP retransmissions), larger buffers will also contribute to buffer bloat latency. 
In our testbed evaluation of AoI for feasible DPDK configurations (section \ref{sec:results}), we will see that age can increase with offered load because the update packets become stale while queued in the ring. Additionally, we will observe how  DPDK employing batch processing to increase packet throughout penalizes the timeliness of update packets.

\subsection{Impact of synchronization primitives}
A conventional packet forwarder is a shared memory multiprocessor machine that maintains a Forwarding Information Base (FIB), a shared memory data structure accessed by both readers and writers. 
For shared memory systems, synchronization primitives allow multiple threads (readers and writers) to execute concurrently and ensure that the results of reading and writing are predictable. 
The literature on synchronization techniques focuses mostly on the scalability, algorithm, implementation, and throughput performance 
in the critical sections\footnote{Formally, a critical section is a protected code segment of the shared resource that is protected against multiple concurrent accesses.} \cite{gramoli2015, davidEPFLasynchronized, clements2012scalable}. 
However, the impact of synchronization primitives on network performance is 
an under-studied problem.


The FIB is typically implemented as a concurrent hash table in which the contention between readers and writers is usually handled either by using a lock-based primitive such as  Readers-Writer lock (RWL) \cite{courtois1971}, or a lock-less primitive such as Read-Copy-Update (RCU) \cite{MckenneyRCU2001, kernelRCU}.
Both RCU and RWL-based data structures have been bench-marked using stand-alone stress tests \cite{kokologiannakisstateless, Kokologiannakis2017, liang2018verification, tassarotti2015verifying, dice2019bravo, passiverwl, nir2013numarwl}, but, when used and implemented in the networking stack, their performance raises many concerns and questions \cite{chronos, shao-sdn-nsdi}. For example, authors in \cite{chronos} studied RCU and RWL and argued that lock contention in a Memcached application accounts for around a third of the overall kernel overhead and significantly contributes to delay and latency variation.

In the context of timeliness, the age performance of RWL and RCU needs to be better understood, especially when these are used in fundamental networking data structures.
The aim of this study is, thus, to understand and analyze the impact of RCU and RWL on the timely updating of shared memory and how this, in turn, affects timely routing of information updates, which is discussed in detail in section \ref{sec:networkforwarding}.
To our knowledge, this is the first quantitative experimental  study of the two widely used synchronization primitives concerning the AoI performance metric.

In section \ref{sec:results}, we show that our AoI experimental results are consistent with the literature that RCU takes advantage of its light read-side primitives to generally outperform RWL.  However, at low packet sending rates, this difference is negligible. The caveat of using RCU, however, is that each write makes a copy of the shared-object,  which means that the memory footprint of the code is larger and also requires a complex garbage collection mechanism \cite{AjitMvrlu}.





\section{Overview of Synchronization Primitives}
Readers-Writer Lock (RWL) is a synchronization primitive that enforces mutual exclusion between readers and writers; multiple readers are allowed to read the shared data structure concurrently, while a writer requires exclusive access or a ``lock''\footnote{In shared-memory multiprocessor architectures, a lock is a mechanism that restricts the access to a shared data structure among multiple processors} to that data structure.
Most RWL implementations are  \textit{write preferring} \cite{courtois1971} i.e., once the writer starts waiting in a queue to acquire the lock, the RWL mechanism prevents new readers from acquiring the lock. The writer's acquisition of the lock occurs once all readers already holding the read lock have finished reading. During the write lock, new read lock requests are queued until the writer has released its lock. 
Note that the RWL protected critical section allows multiple readers to read that critical section concurrently but written exclusively.

Replacing conventional locking techniques, Read-Copy-Update (RCU) is a synchronization primitive that allows concurrent forward progress for both writers and readers \cite{MckenneyRCU2001}. RCU can be broadly described in two steps \cite{kernelRCU}: 
\begin{enumerate*}
     \item To publish a fresher version of a data item, 
the writer creates a copy of the RCU protected data item, modifies this copy,
     and atomically replaces the old reference with a reference to this newer version.
     This publishing process runs concurrently with ongoing read processes that continue to read the old copy/version using the old reference. However, subsequent read requests 
     obtain the freshest   version. 
    \item Since some readers in progress hold reference to ``stale'' data, the system defers memory reclamation of old data until each in-progress reader  has finished executing its read-side critical section. 
\end{enumerate*}
Therefore,  RCU typically maintains multiple versions of a data item that are concurrently being read. 
Note that an RCU read-side critical section guarantees that a read-locked copy of the resource is not reclaimable for the full duration of that critical section.


\section{Timely Update Forwarding}
\label{sec:networkforwarding}
To demonstrate the effects of packet buffering, batch processing, and FIB concurrency constructs on 
timeliness,
we consider an example of packet forwarding in a mobile user environment, as shown in Fig.~\ref{fig:experimentconcept}. An application server in the network is sending ``app updates'' regarding a process of interest to a mobile terminal. The application sends its update packets to a forwarding node in the network. This forwarder maintains a Forwarder Information Base (FIB) that tracks the location (i.e. point of attachment network address) of the mobile terminal. At the forwarder, app updates are addressed using the FIB and forwarded to the mobile terminal.

In this system, we will track two update age processes:
\begin{enumerate*} \item the age $\age(t)$ of app updates at the mobile, \item the age $\hat{\age}(t)$ of mobile user ``location updates'' written in the FIB. \end{enumerate*}
These age processes are coupled through the FIB -- {\em the location updates  written to the FIB and the app updates that induce client requests to read the FIB  both contend for access to the FIB.} Moreover, if the FIB holds an outdated (i.e. wrong) address, the misaddressed app updates are lost in transit and such packet losses will increase the age of app updates at the mobile terminal. We will see with the experiment results that misaddressed packets can arise in RCU if the reader reads the FIB while the write of a fresh  location update is in progress. On the other hand, misaddressed packets occur in RWL when a read lock prevents the writer from writing a fresh location update.

\section{Experiment Design and Testbed}
\label{sec:setup}

\begin{figure}[t]
    \centering
\includegraphics[width=0.49\textwidth]{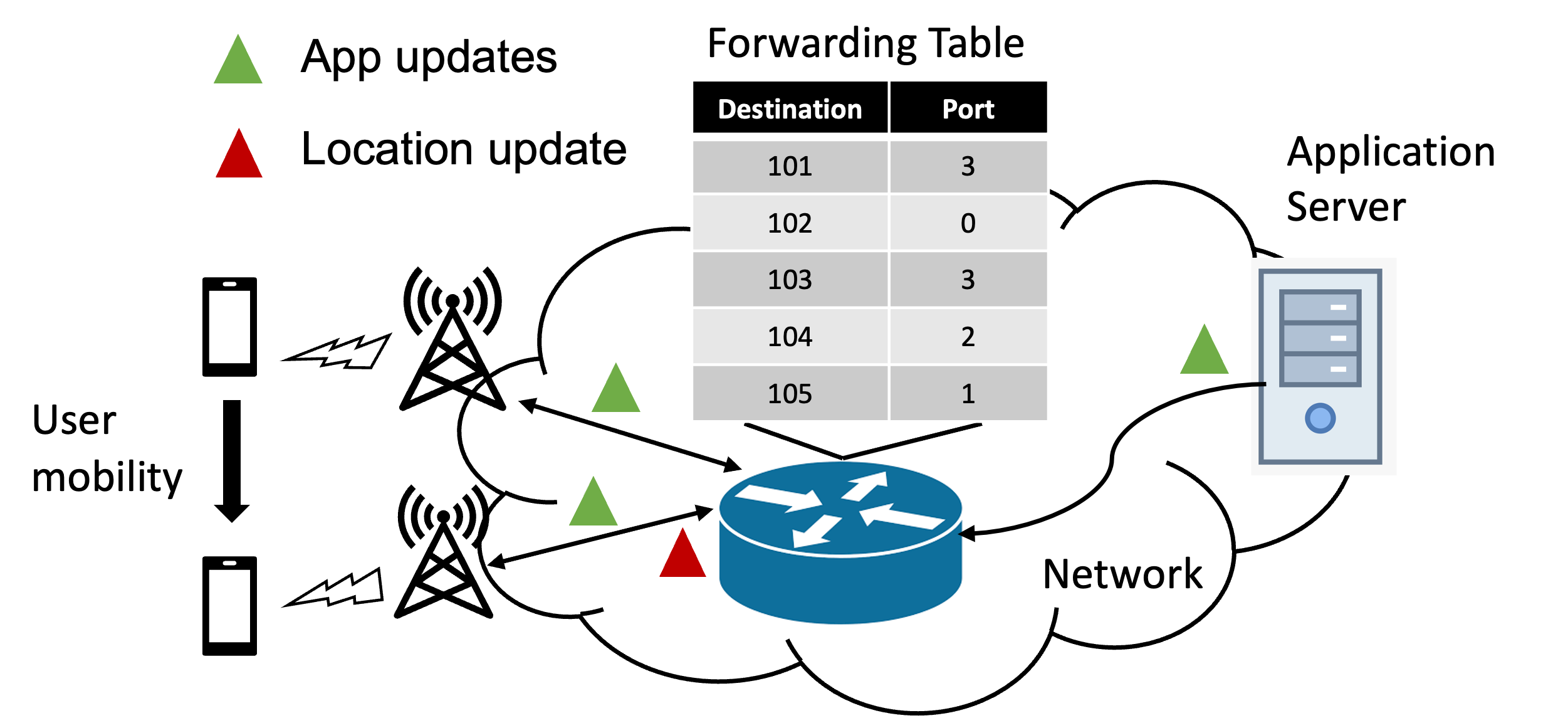}
    \caption{Update  forwarding for mobile users}
\label{fig:experimentconcept}
    \vspace{-5mm}
\end{figure}

This section describes the experimental setup used to evaluate the layer 2 packet forwarding application depicted in Fig.~\ref{fig:experimentconcept}.
%
We note that AoI evaluation 
requires time-based computations across machines. Since the age of an update packet is based on a timestamp inserted by the sender,  calculation of the age of an update at the receiver requires synchronized clocks at the sender and receiver. However, the accuracy of the NTP protocol supported by the testbed 
is around 1ms,  which is too coarse to measure delays on a microsecond scale.\footnote{We note this is merely a limitation of our testbed. Sub-microsecond timing accuracy is feasible, although not generally used in network routers \cite{GuoCrossley2017}.}
Our experimental workaround is to place the sender and receiver functionality on the same machine.

\begin{figure}[t]
    \centering
\includegraphics[width = 0.4\textwidth, height = 2.5in]{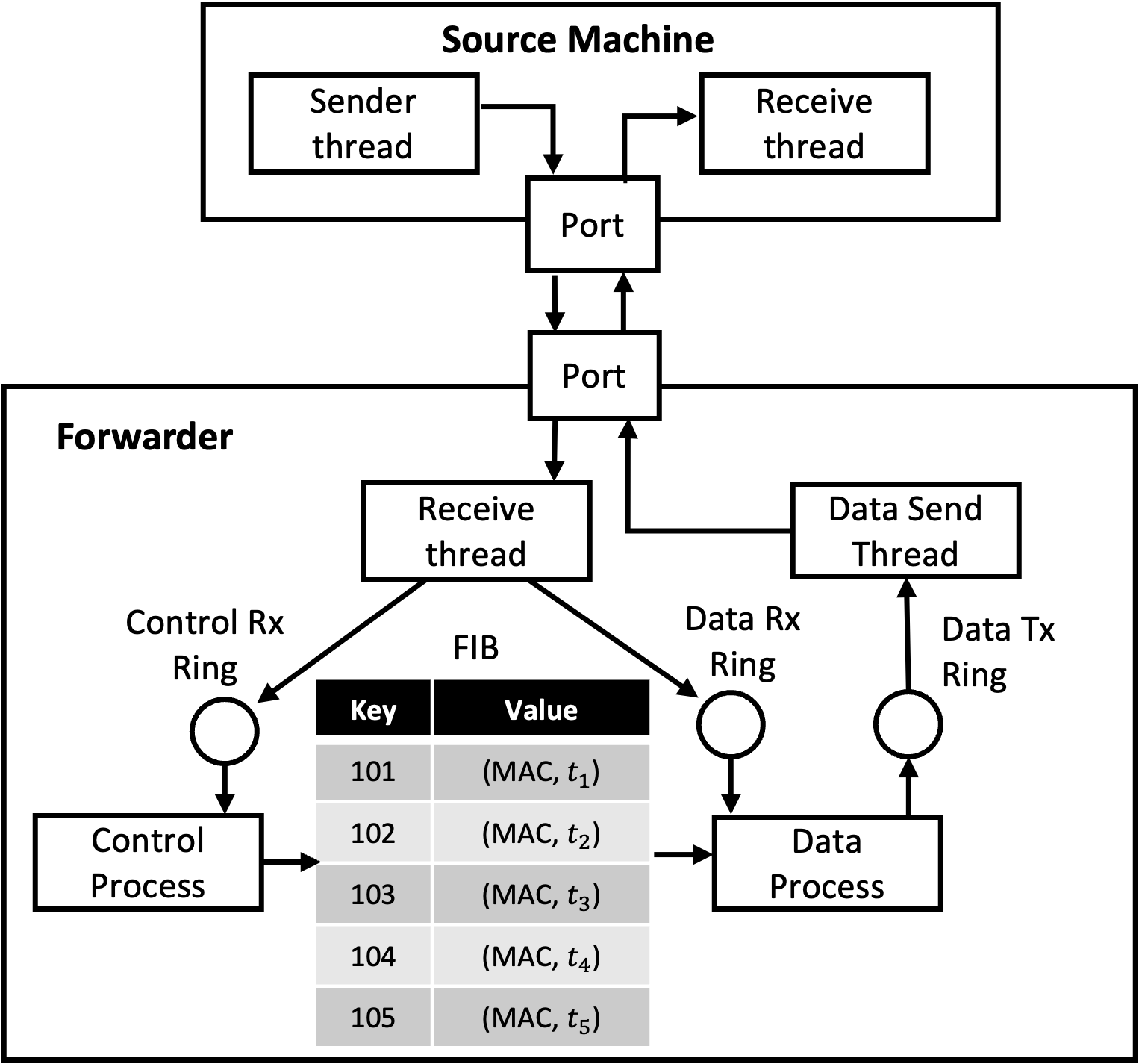}
 \caption{Packet forwarding testbed: The Source machine emulates the app update sender and receivers as well as their location update senders. In the Forwarder, the FIB stores the key-value pair as user ID (101, 102, \ldots) and address tuple while the control and data processes  contend for FIB access.}
\label{fig:routingexp}
    \vspace{-5mm}
\end{figure}

Fig.~\ref{fig:routingexp} shows a block diagram of a testbed architecture with two machines (Source and Forwarder) that implements the system shown in Figure~\ref{fig:experimentconcept}. 
The sender thread in the Source acts as the application server that sends time-stamped app updates (data packets) to multiple mobile users.
This sender thread also 
emulates mobile user movement by sending time-stamped location updates (control packets). Each update carries a user ID indicating a location change for that user. 
The receive thread in the Source acts as the mobile users receiving app update packets. Each time-stamped update carries a user ID that enables the Source to track the app update age process of each user. 
Although the scale of the experiment is small, using only two machines allows us to focus on the primary bottlenecks we have identified: compute bottleneck due to synchronization primitives, and queuing bottleneck at the sender.

At the Forwarder, the FIB is 
implemented
as a hash table for fast lookup. 
The destination user ID acts as a key, and the hash function translates this key into a hash index that points to an {\em address tuple}. This address tuple consists of a MAC address and a timestamp as shown in Fig.~\ref{fig:routingexp}. 
A traditional FIB would store a next hop MAC address for a destination user ID and update this MAC address to reflect a new point of attachment for the mobile user. Since the next hop MAC address (Source machine) is same for all users in our testbed implementation, old and new address entries in the FIB are distinguished by the timestamp in the address tuple. 
Upon receiving a new 
control packet from the Source, the control process in the Forwarder updates the timestamp in the address tuple for the corresponding user ID. 

To ensure reproducibility of the experiments, we use a trace file that consists of rows of type {\em <type, userID>}, where a type indicates whether a packet sent from the Source is a control or data packet.
The order of packet types in the trace is decided by a pseudo-random sequence of coin flips such that the control data ratio (CDR) parameter specifies the ratio of control and data packets.

The user IDs for both data and control packets are selected from a Zipf distribution with exponent $1$ on a set of 1000 user addresses. To eliminate randomness of the packet preparation time,
packets are  read from the trace file and stored in the  memory of the sender thread of the Source before the experiment is run.
The sender thread sends both control and data packets from the same interface due to hardware and software limitations and to enforce a  packet trace sequence. 

At the Forwarder, the receive thread pulls packets from the receive ring of the NIC and moves control and data  packets to their respective Rx rings.
The control process 
retrieves the control packets from the control Rx ring and updates the FIB, thus acting as a {\em writer}.
The data process retrieves data packets from the data Rx ring and, acting as the FIB {\em reader}, addresses each data packet via a lookup of the user ID carried by the packet.
Each data process read of the FIB returns with an address tuple. The corresponding timestamp in the tuple is inserted in the header of the data packet which is then sent to Data Tx Ring. These modified data packets, which represent app updates, are then sent back to the Source.

For each user ID, an address tuple is said to be {\em fresh} if its timestamp is the same as that of the last sent address update (control packet) from the Source. That is, an address is fresh if the mobile user has not sent a subsequent location update.
In our experiments, the freshness of the address determines the status of a data packet received by receive thread. Specifically, received data packets with a fresh address are classified as correctly received app updates and serve to reduce the app update age of the corresponding mobile user.   On the other hand, a received data packet with an address that is not fresh is classified as misaddressed and
regarded as lost in transit. 

We note that buffer overflow events (i.e. packet drops) occur at the control (data) Rx ring when the 
control (data) process at the Forwarder fails to keep up with its incoming packet stream. 
A dropped control packet signifies that a user movement has not been updated in the FIB, resulting in misaddressed data packets. A dropped data packet indicates that an app update has not been received at the mobile user, hence increasing that user's app update age.
\section{Testbed Results}
\label{sec:results}
Experiments are executed on the COSMOS experimental networking testbed \cite{Cosmos}.
We employ DPDK, 
a set of data plane libraries and network interface card (NIC) drivers to support fast packet processing in user space \cite{dpdk}.
The machines use Intel(R) Xeon(R) Gold 6126 CPU @ 2.60GHz (24 cores, hyper-threading and turbo-boost turned off) with 192GB RAM evenly distributed on 2 NUMA nodes. Each thread in our experiment is pinned to each core in a single NUMA node which in turn is pinned to a single socket.
A Mellanox ConnectX-4 Lx 25GbE network interface card is connected to NUMA node 0.
We run our program on Ubuntu 18.04.6 LTS, and DPDK 21.08. The Source and Forwarder are connected via a dedicated switch with 25Gbps Ethernet links.



\begin{table}[t]
    \centering
\resizebox{\columnwidth}{!}{%
    \begin{tabular}{|p{0.12\textwidth} | p{0.07\textwidth} | p{0.07\textwidth} | p{0.04\textwidth} | p{0.08\textwidth}|}
    \hline
    {\bf{Experiment}} & {\bf{Data pkts}} & {\bf{Ctrl pkts}} & {\bf{Users}}&{\bf{FIB lookup}}\\
         \hline
         Baseline &  47996440 & 0 & 1 &No\\ \hline
         Routing, CDR 0.01 
         & 39996600 & 400000 & 1000 & Yes \\ \hline
         Routing, CDR 0.1
         & 39996600 & 3999800  & 1000 & Yes\\
         \hline
    \end{tabular}
    }
    \caption{All experiments share the following DPDK configurations: (1) Source-Tx burst size 32, Tx ring size 64, Rx burst size 64, Rx ring size 4096. (2) Forwarder-Tx/Rx burst size 64, Tx/Rx ring size 4096.}
    \label{tab:exp-params}
    \vspace{-5mm}
\end{table}

Table \ref{tab:exp-params} summarizes three 
sets of experiments performed in this study. 
In the testbed,  data and control packets are both $60$ bytes long and the Source   can pump  these packets at a maximum rate of $R_{\max}=14$~Mpps (million packets per second). In our experiments, our results are shown as a function of the offered load $R$~pps. Specifically, the sender thread in the Source feeds the  pre-prepared packet trace using a token bucket rate control mechanism and a maximum data burst size of $B$ packets (in our experiments, $B = 32$). At time $t_0=0$, the bucket is initialized with $N_0=0$ tokens and 
tokens then accumulate at a rate of $R$ tokens/s. The sender thread requests to place packets on the Tx ring of the Source NIC at times $t_0,t_1,\ldots$ such that at time $t_i$ with $N_i$ tokens, the sender thread calls the {\em eth\_tx\_burst} function to offer $K_i=\min(B,N_i)$ packets to the NIC. At time $t_{i+1}$,  {\em eth\_tx\_burst} returns that a {\em batch} of $L_i$ packets were admitted to the NIC and thus 
$N_{i+1}=N_i-L_i+(t_{i+1}-t_i)R$ tokens are available to repeat this process until the entire packet trace is admitted to the Source NIC.


Because no packets are dropped at the Source Tx ring, $L_i<K_i$ indicates that {\em eth\_tx\_burst} call filled the ring. Also note that while the {\em eth\_tx\_burst} execution time $\tau_i=t_{i+1}-t_i$ is random\footnote{We observe that $\tau_i$ and $L_i$ appear to be weakly but positively correlated.}, the process self-adjusts to offer packets at rate $R$ pps for all $R<R_{\max}$. 
Finally, we note that hardware limitations dictate that all $L_i$ packets in a batch are recorded with the same timestamp $t_i$, and this timestamp is inserted by the CPU. 
While this sending process 
is not the usual Poisson update process in analytical studies, it does offer a repeatable characterization of AoI performance under variable offered load.

We note that high-speed packet IO such as DPDK use large batches by default, leading to a trade-off between bursty high throughput and precise packet generation. Cases where users might not require a bursty traffic, such as in generating Poisson stream of packets are difficult to emulate reliably, especially at high sending rates. Packet data cannot be directly sent to the NIC, but can be placed in a DMA memory region and retrieved asynchronously by the NIC, causing unwanted jitter \cite{Gallenmuller-mindthegap}. Further, calling the eth\_tx\_burst() function to place packets on the output ring takes some random time to return, which also depends on the number of packets offered. 

For Poisson packet arrivals, a pure software approach would wait for pre-configured pseudorandom times between sending individual packets. Implementing a close approximation to exponential inter-arrivals is not a problem at low packet rates where the aforementioned random system delays are negligible compared to inter-packet times. However, Poisson arrival emulation becomes difficult at rates approaching the limitations of the testbed hardware and the software framework running on it. When packet delays from the system are a significant fraction of the average inter-arrival time, the precision of Poisson traffic pattern remains a concern.

\subsubsection*{Baseline Experiment}
\begin{figure}[t]
    \begin{tabular}{c}
    \includegraphics[width = 0.35\textwidth,height=1.5in
    ]{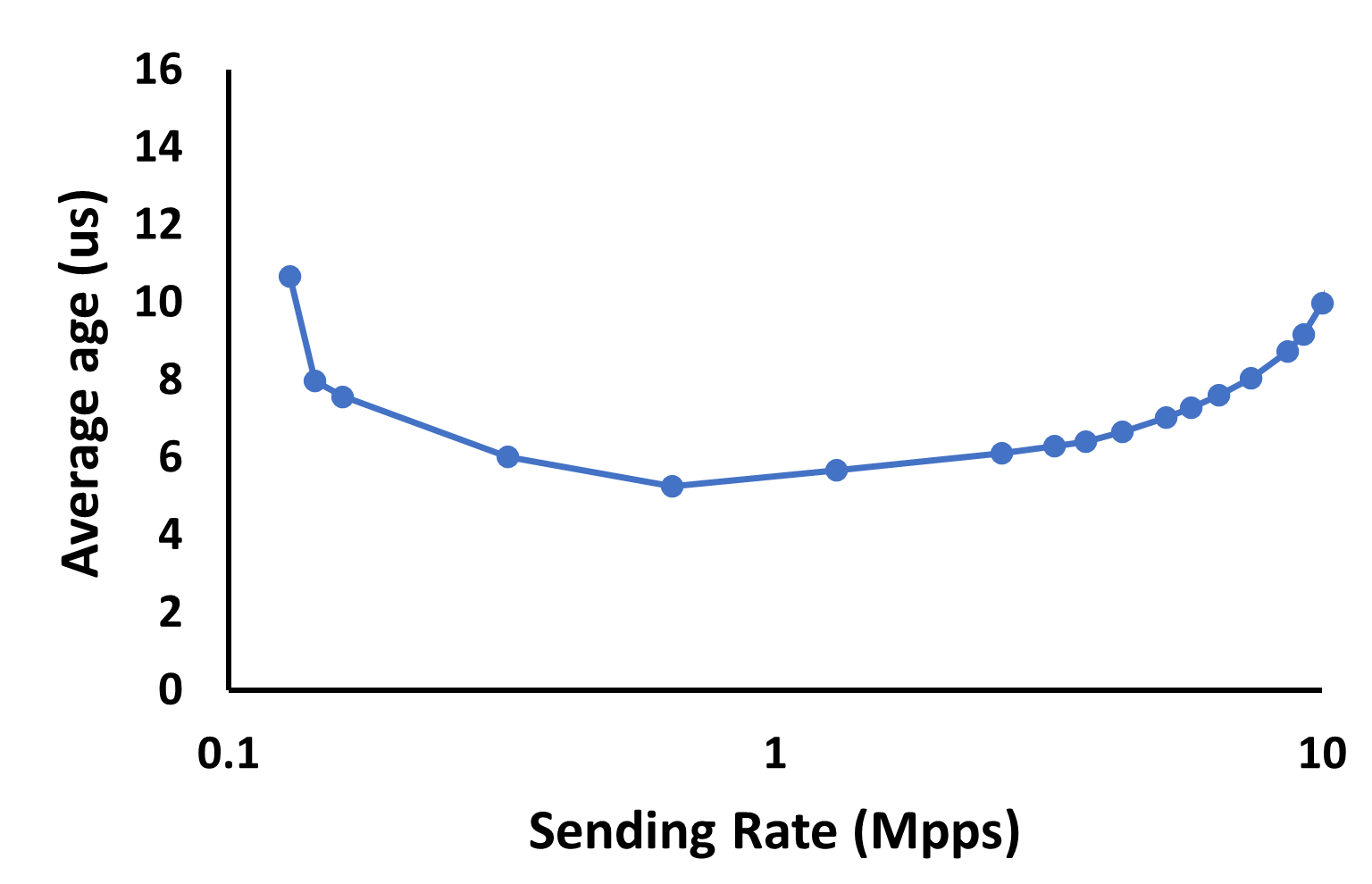} \\
    {\bf{(a)}} \\
    \includegraphics[width=0.35\textwidth,height=1.5in
    ]{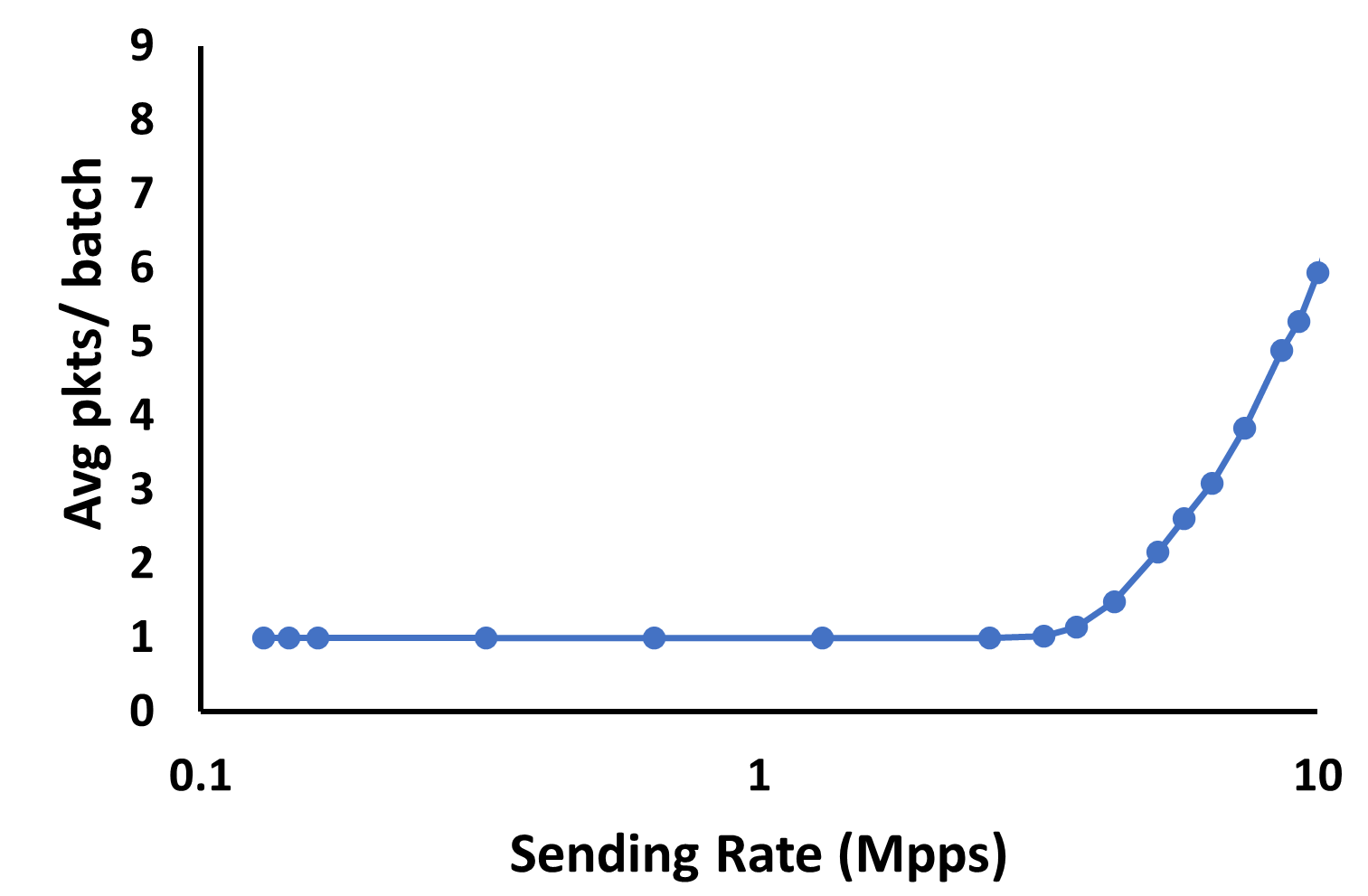} \\
    {\bf{(b)}}
    \end{tabular} 
    \vspace{-2mm}
    \caption{Baseline experiment}
    \vspace{-2mm}
    \label{fig:baseline}
\end{figure}

To understand the rate control mechanism, we performed a baseline experiment with a single immobile user.  No control packets were sent and the FIB addressing mechanism at the Forwarder was bypassed  so that all packets were immediately sent back to the Source. In Fig.~\ref{fig:baseline}(a), we see that for $R<1$ Mpps, the average age initially declines with $R$ (as expected) because updates become less infrequent. However, perhaps unexpectedly, we see that as $R$ becomes large, the average age grows. This is a consequence of rate control.  Fig.~\ref{fig:baseline}(b) presents the average batch size for all nonzero batch sizes. For $R< 4$~Mpps, tokens accumulate slowly and each {\em eth\_tx\_burst} call offers either zero or one packet. Thus each admitted batch has only a single packet. However, for $R>4$~Mpps, the average batch size grows with $R$, and the growing average age reflects the input queueing induced by the batch admission procedure.  In short, 
processing more packets in a batch increases throughput, 
but this is not necessarily favorable to timeliness.

\begin{figure*}[t]
\begin{tabular*}{\textwidth}{c  @{}  c @{}  c }
    \includegraphics[width=0.32\textwidth, height=1.5in]{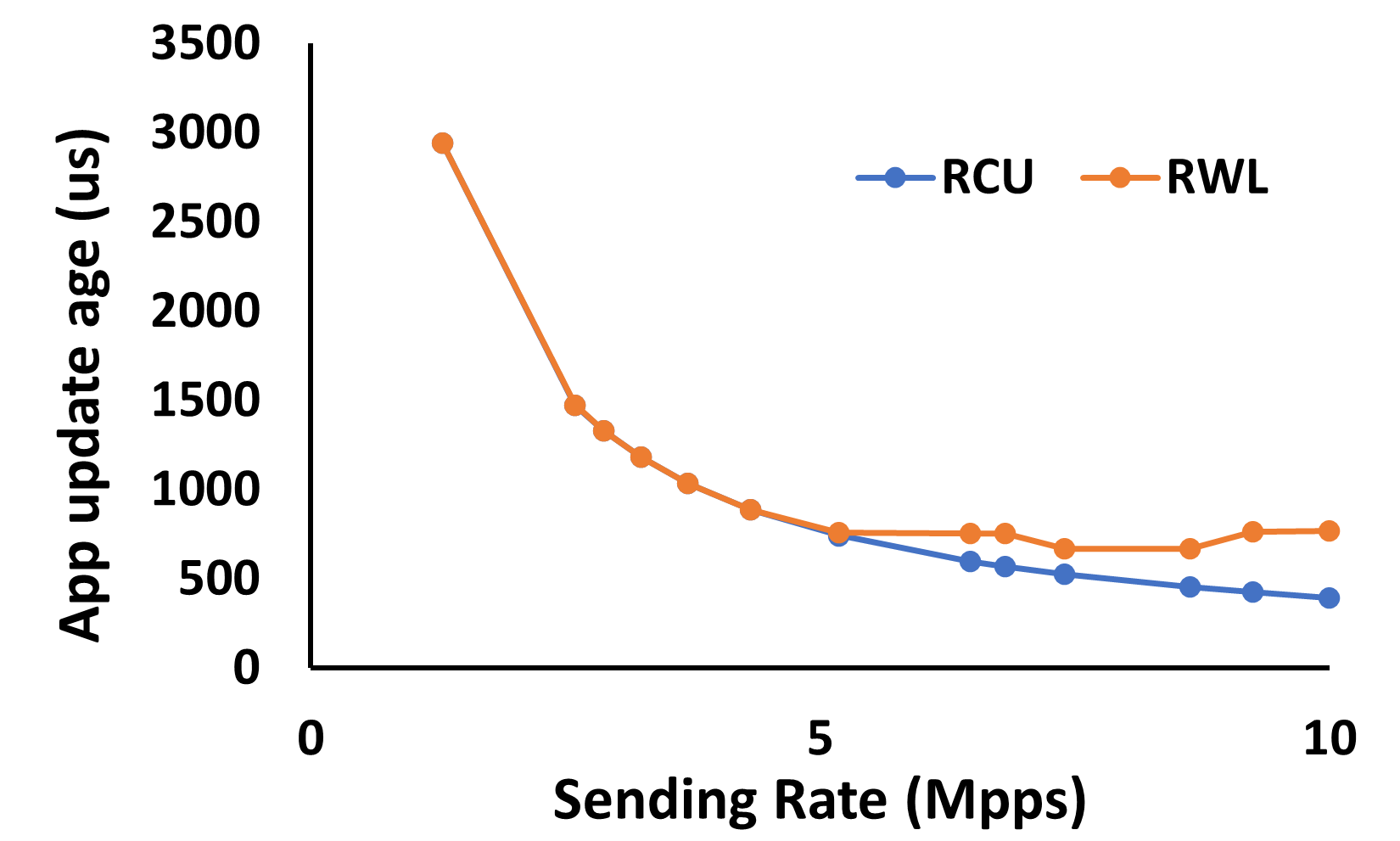}&
    \includegraphics[width=0.32\textwidth, height=1.5in]{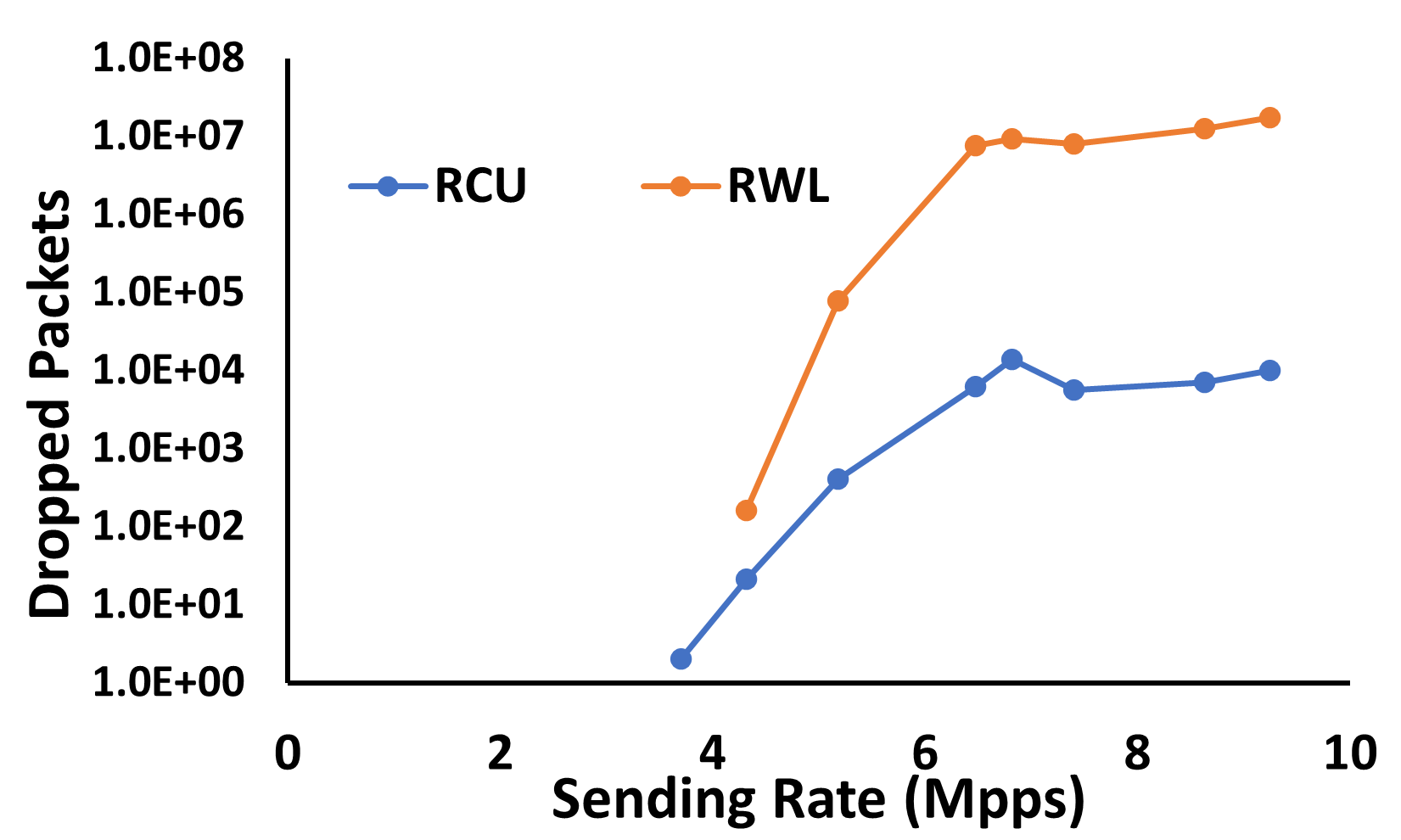}&
    \includegraphics[width=0.32\textwidth, height=1.5in]{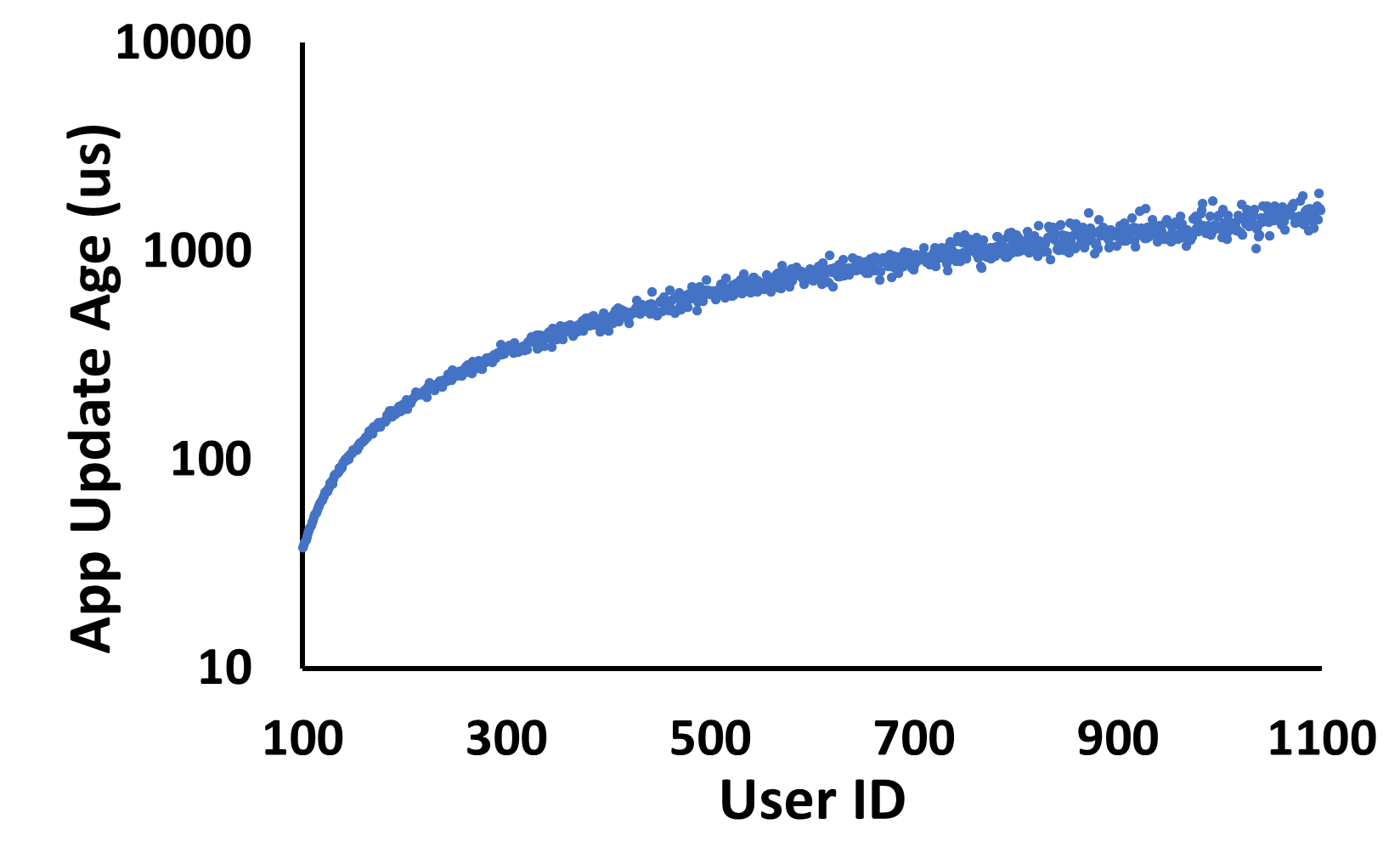}
    \\
    \bf{(a)} $\text{CDR} = \mathbf{0.01}$  & \bf{(b)} $\text{CDR} = \mathbf{0.01}$ & \bf{(c)}
    \\
    \includegraphics[width=0.32\textwidth, height=1.5in]{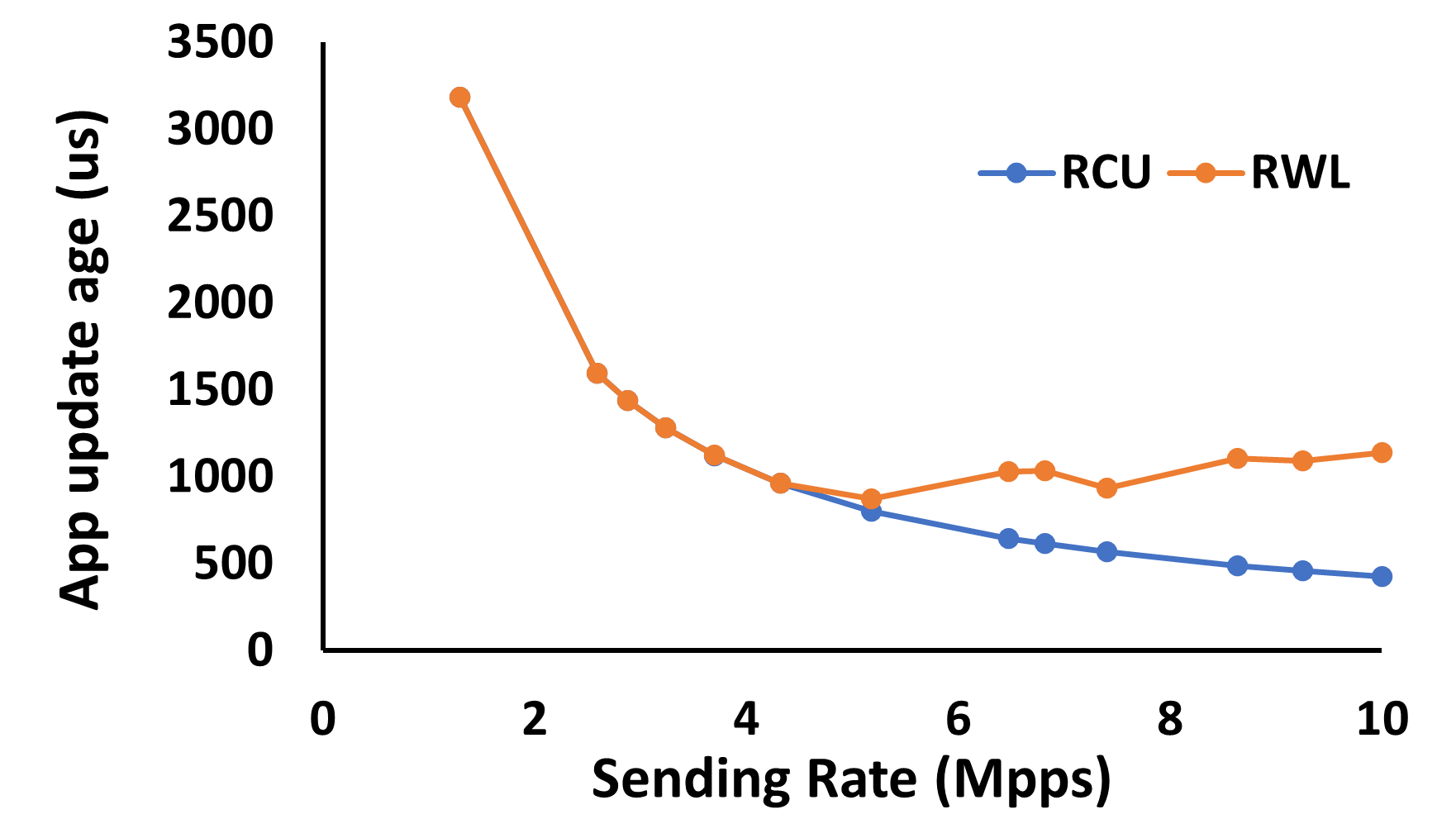}&
    \includegraphics[width=0.32\textwidth, height=1.5in]{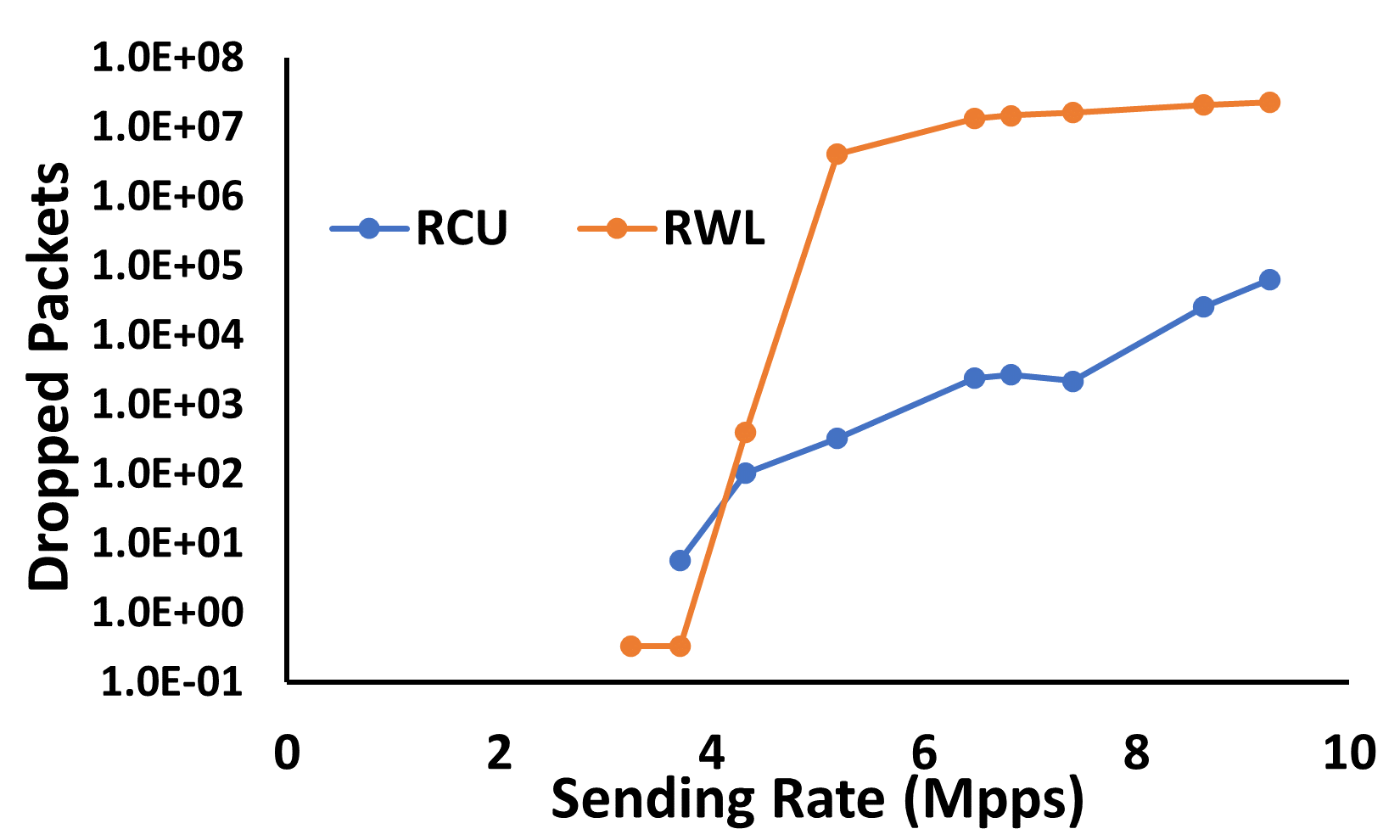}&
    \includegraphics[width=0.32\textwidth, height=1.5in]{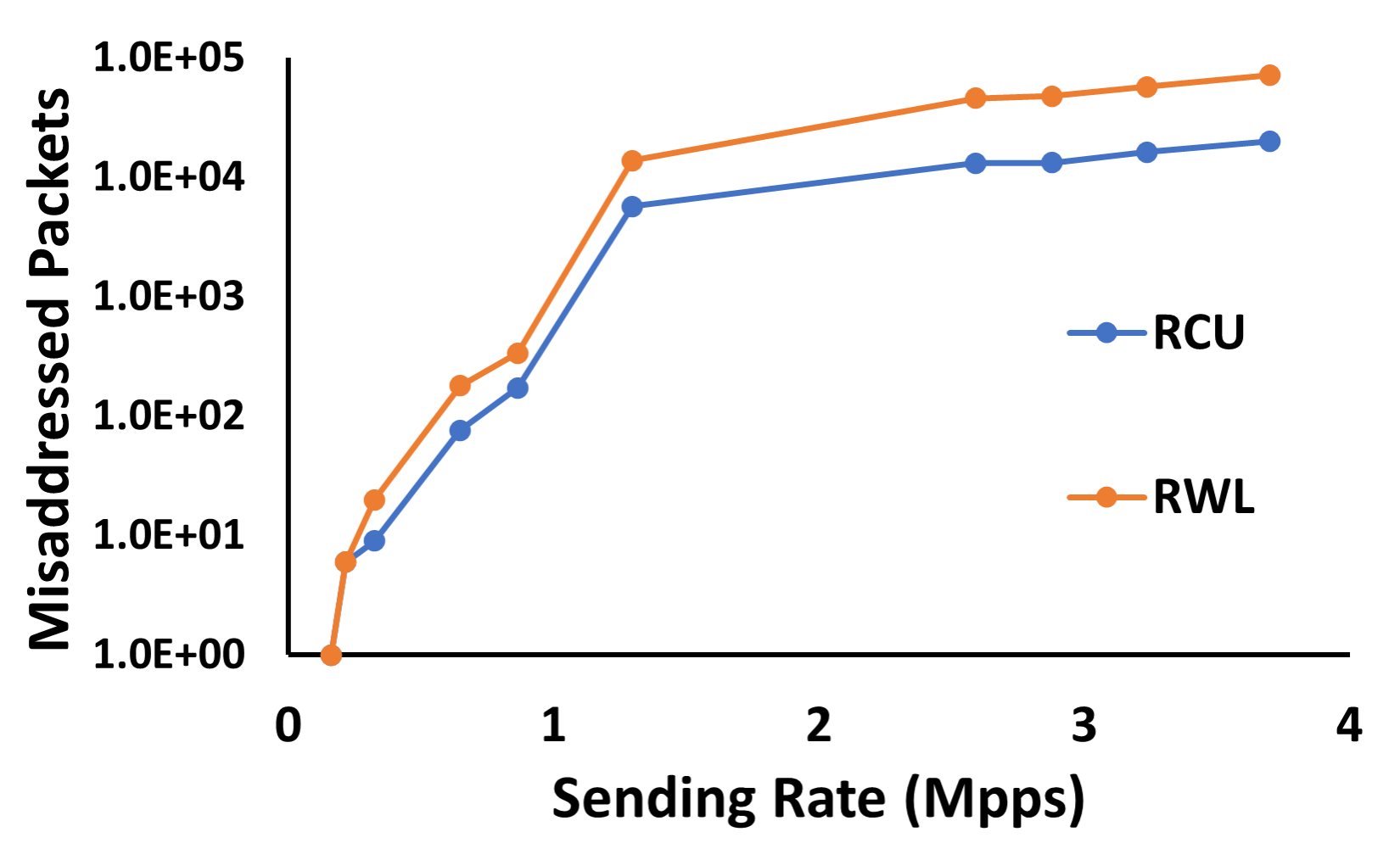}
    \\
    \bf{(d)} $\text{CDR} = \mathbf{0.1}$ & \bf{(e)}  $\text{CDR} = \mathbf{0.1}$ & \bf{(f)} $\text{CDR} = \mathbf{0.1}$ \\
    [-5pt]
\end{tabular*}
\caption{Results from packet forwarding testbed (Fig.~\ref{fig:routingexp}) with control/data Rx ring sizes $=\mathbf{64}$, data tx ring size $=\mathbf{1024}$. The plots depict age performance of Read-Copy-Update (RCU) and Readers-Writer Lock (RWL) as a function of the sending rate $R$ Mpps.  
}
\label{fig:results}
\end{figure*}

\subsubsection*{Routing Experiments}
We now examine  the effect of the FIB access mechanism (RCU or RWL) on the average age of app updates for a set of 1000 mobile users.  
In these experiments, the packet stream, with sending rates ranging from $R=1$ to $R=10$~Mpps, represents the aggregated updating processes of all 1000 users. Compared to the prior baseline experiment with one user, the packet sending rates associated with a particular user are scaled by the Zipf distribution probabilities. Average update ages of 1-10~$\mu$s  seen in the one-user baseline experiment become 5-10~ms in averaging over all 1000 users; Fig.~\ref{fig:results}(c) shows the average app update age for each user for $R = 10$ Mpps under the RWL construct.

A subset of experimental results is shown in Fig.~\ref{fig:results} for $\text{CDR}=0.01$ and $\text{CDR}=0.1$. We note that increasing the CDR stresses the system in two ways. First, it increases FIB access contention between readers and the writer. Second, increasing the CDR corresponds to increased change in mobile users location, thus, increasing the likelihood that app updates are misaddressed.


In Figs.~\ref{fig:results}(a)-(b) we see for $\text{CDR}=0.01$ that the Forwarder performs reasonably well at all packet rates.
Fig.~\ref{fig:results}(a) shows the average app update age generally decreasing with the packet sending rate, with RCU 
outperforming RWL at high packet rates. 
This is consistent with relatively low rates of dropped 
data packets in Fig.~\ref{fig:results}(b).  
However, at higher sending rates, mutual exclusion between reads and write in RWL slows the data process to handle the incoming packets on the data Rx ring of the forwarder, resulting in increased data packet drops at higher sending rates, see Fig.~\ref{fig:results}(b). 

For $\text{CDR}=0.1$, Figs.~\ref{fig:results}(d)-(f) reflect the increased stress of a high CDR. In particular, the results reinforce the fact that neither RCU nor RWL is good in update-heavy scenarios as RCU writes are heavy and RWL enforces mutual exclusion.
These mechanisms effectively slow the control processing so that the control Rx ring is quickly filled at a higher control packet rates
and the dropping rate of  control packets is high. As a consequence, the FIB is updated with stale control updates, increasing the rate of misaddressed data packets, as seen in 
Fig.~\ref{fig:results}(f), and increasing the app update age in Fig.~\ref{fig:results}(d), as compared to $\text{CDR} = 0.01$. The number of misaddressed packets was calculated based on the total number of received packets on Source machine that carried timestamp older than the last sent control timestamp for that user ID. Note that in Fig.~\ref{fig:results}(f), we plot misaddressed packets only for $R < 4$ Mpps. This is because at $R >4$ Mpps, we observed significant data packet drops at the Forwarder, resulting in fewer received packets at the Sender. Calculating misaddressed packets based on these fewer packets gave us statistically incorrect number. 

RCU is read-friendly, as is evident in Fig.~\ref{fig:results}(e), with fewer data packet drops. By contrast, RWL reads are frequently locked out at higher write request rates, 
 This slows data packet processing under RWL and increases the data packet drops at the data Rx ring. Consequently, Fig.~\ref{fig:results}(d) shows that the average age under RWL follows the classic pattern of updating systems: age initially decreases with the update rate but eventually increases as the system becomes congested \cite{Kaul-YG-infocom2012}. In short, updating  should be fast, but not too fast. 

Lastly, we note that the average location update age at the FIB decreases with sending rate for both $\text{CDR} = 0.01$ and $\text{CDR} = 0.1$ and that RCU
and RWL all have similar impact on location update age. However, space constraints precluded presentation of 
numerical results. 

\section{Conclusion}
In this work, we designed and implemented a DPDK-based packet forwarding experiment. 
We quantitatively evaluated the performance of the readers-writer lock (RWL) and (lock-less) read-copy-update (RCU) synchronization primitives, in terms of the Age-of-Information (AoI) performance metric. Even in a  relatively simple one-forwarder system, this initial study revealed complex interactions between FIB synchronization mechanisms and packet queueing. This work highlights how more work is needed on optimizing packet processing frameworks such as DPDK for updating systems. 

\bibliographystyle{IEEEtran}
\bibliography{refs,AOI-2020-03}

\begin{thebibliography}{10}
\providecommand{\url}[1]{#1}
\csname url@samestyle\endcsname
\providecommand{\newblock}{\relax}
\providecommand{\bibinfo}[2]{#2}
\providecommand{\BIBentrySTDinterwordspacing}{\spaceskip=0pt\relax}
\providecommand{\BIBentryALTinterwordstretchfactor}{4}
\providecommand{\BIBentryALTinterwordspacing}{\spaceskip=\fontdimen2\font plus
\BIBentryALTinterwordstretchfactor\fontdimen3\font minus
  \fontdimen4\font\relax}
\providecommand{\BIBforeignlanguage}[2]{{%
\expandafter\ifx\csname l@#1\endcsname\relax
\typeout{** WARNING: IEEEtran.bst: No hyphenation pattern has been}%
\typeout{** loaded for the language `#1'. Using the pattern for}%
\typeout{** the default language instead.}%
\else
\language=\csname l@#1\endcsname
\fi
#2}}
\providecommand{\BIBdecl}{\relax}
\BIBdecl

\bibitem{dtremotesurgery}
H.~Laaki, Y.~Miche, and K.~Tammi, ``Prototyping a digital twin for real time
  remote control over mobile networks: Application of remote surgery,''
  \emph{IEEE Access}, vol.~7, pp. 20\,325--20\,336, 2019.

\bibitem{surgendoscopy}
R.~K. Orosco, B.~Lurie, T.~Matsuzaki, E.~K. Funk, V.~Divi, F.~C. Holsinger,
  S.~Hong, F.~Richter, N.~Das, and M.~Yip, ``Compensatory motion scaling for
  time-delayed robotic surgery,'' \emph{Surgical endoscopy}, vol.~35, pp.
  2613--2618, 2021.

\bibitem{belay2014ix}
A.~Belay, G.~Prekas, A.~Klimovic, S.~Grossman, C.~Kozyrakis, and E.~Bugnion,
  ``$\{$IX$\}$: a protected dataplane operating system for high throughput and
  low latency,'' in \emph{11th USENIX Symposium on Operating Systems Design and
  Implementation (OSDI 14)}, 2014, pp. 49--65.

\bibitem{pontarelli-flowblaze}
\BIBentryALTinterwordspacing
S.~Pontarelli, R.~Bifulco, M.~Bonola, C.~Cascone, M.~Spaziani, V.~Bruschi,
  D.~Sanvito, G.~Siracusano, A.~Capone, M.~Honda, F.~Huici, and G.~Siracusano,
  ``{FlowBlaze}: Stateful packet processing in hardware,'' in \emph{16th USENIX
  Symposium on Networked Systems Design and Implementation (NSDI 19)}.\hskip
  1em plus 0.5em minus 0.4em\relax Boston, MA: USENIX Association, Feb. 2019,
  pp. 531--548. [Online]. Available:
  \url{https://www.usenix.org/conference/nsdi19/presentation/pontarelli}
\BIBentrySTDinterwordspacing

\bibitem{kosta2017age}
A.~Kosta, N.~Pappas, and V.~Angelakis, ``Age of information: A new concept,
  metric, and tool,'' \emph{Foundations and Trends in Networking}, vol.~12,
  no.~3, pp. 162--259, 2017.

\bibitem{Yates-SBKMU-2021jsac-survey}
R.~D. Yates, Y.~Sun, D.~R. Brown, S.~K. Kaul, E.~Modiano, and S.~Ulukus, ``Age
  of information: An introduction and survey,'' \emph{IEEE Journal on Selected
  Areas in Communications}, vol.~39, no.~5, pp. 1183--1210, 2021.

\bibitem{Kadota2020age}
I.~Kadota, M.~S. Rahman, and E.~Modiano, ``Age of information in wireless
  networks: From theory to implementation,'' in \emph{Proceedings of the 26th
  Annual International Conference on Mobile Computing and Networking}, 2020,
  pp. 1--3.

\bibitem{Kadota2021wifresh}
------, ``Wifresh: Age-of-information from theory to implementation,'' in
  \emph{2021 International Conference on Computer Communications and Networks
  (ICCCN)}.\hskip 1em plus 0.5em minus 0.4em\relax IEEE, 2021, pp. 1--11.

\bibitem{Shreedar-KY-wowmom2019}
T.~Shreedhar, S.~K. Kaul, and R.~D. Yates, ``An age control transport protocol
  for delivering fresh updates in the internet-of-things,'' in \emph{20th IEEE
  International Symposium on A World of Wireless, Mobile and Multimedia
  Networks {(WoWMoM)}}, 2019.

\bibitem{dpdk}
``{Data Plane Development Kit},'' [Online]. Available from:
  \url{https://www.dpdk.org/}, 2021.

\bibitem{Kaul-YG-infocom2012}
S.~Kaul, R.~Yates, and M.~Gruteser, ``Real-time status: How often should one
  update?'' in \emph{Proc. IEEE INFOCOM}, March 2012, pp. 2731--2735.

\bibitem{Costa-CE-IT2016management}
M.~Costa, M.~Codreanu, and A.~Ephremides, ``On the age of information in status
  update systems with packet management,'' \emph{IEEE Trans. Info. Theory},
  vol.~62, no.~4, pp. 1897--1910, April 2016.

\bibitem{gramoli2015}
\BIBentryALTinterwordspacing
V.~Gramoli, ``More than you ever wanted to know about synchronization:
  Synchrobench, measuring the impact of the synchronization on concurrent
  algorithms,'' in \emph{Proceedings of the 20th ACM SIGPLAN Symposium on
  Principles and Practice of Parallel Programming}, ser. PPoPP 2015.\hskip 1em
  plus 0.5em minus 0.4em\relax New York, NY, USA: Association for Computing
  Machinery, 2015, pp. 1--10. [Online]. Available:
  \url{https://doi.org/10.1145/2688500.2688501}
\BIBentrySTDinterwordspacing

\bibitem{davidEPFLasynchronized}
T.~David, R.~Guerraoui, and V.~Trigonakis, ``Asynchronized concurrency: The
  secret to scaling concurrent search data structures,'' \emph{ACM SIGARCH
  Computer Architecture News}, vol.~43, no.~1, pp. 631--644, 2015.

\bibitem{clements2012scalable}
A.~T. Clements, M.~F. Kaashoek, and N.~Zeldovich, ``Scalable address spaces
  using rcu balanced trees,'' \emph{ACM SIGPLAN Notices}, vol.~47, no.~4, pp.
  199--210, 2012.

\bibitem{courtois1971}
P.-J. Courtois, F.~Heymans, and D.~L. Parnas, ``Concurrent control with
  ``readers'' and ``writers'','' \emph{Communications of the ACM}, vol.~14,
  no.~10, pp. 667--668, 1971.

\bibitem{MckenneyRCU2001}
P.~E. Mckenney, J.~Appavoo, A.~Kleen, O.~Krieger, O.~Krieger, R.~Russell,
  D.~Sarma, and M.~Soni, ``Read-copy update,'' in \emph{In Ottawa Linux
  Symposium}, 2001, pp. 338--367.

\bibitem{kernelRCU}
P.~E. McKenney, [Online]. Available from:
  \url{https://www.kernel.org/doc/html/latest/RCU/whatisRCU.html}.

\bibitem{kokologiannakisstateless}
M.~Kokologiannakis and K.~Sagonas, ``Stateless model checking of the linux
  kernel's read--copy update (rcu),'' \emph{International Journal on Software
  Tools for Technology Transfer}, vol.~21, no.~3, pp. 287--306, 2019.

\bibitem{Kokologiannakis2017}
\BIBentryALTinterwordspacing
------, ``Stateless model checking of the linux kernel's hierarchical
  read-copy-update (tree rcu),'' in \emph{Proceedings of the 24th ACM SIGSOFT
  International SPIN Symposium on Model Checking of Software}, ser. SPIN
  2017.\hskip 1em plus 0.5em minus 0.4em\relax New York, NY, USA: Association
  for Computing Machinery, 2017, pp. 172--181. [Online]. Available:
  \url{https://doi.org/10.1145/3092282.3092287}
\BIBentrySTDinterwordspacing

\bibitem{liang2018verification}
L.~Liang, P.~E. McKenney, D.~Kroening, and T.~Melham, ``Verification of
  tree-based hierarchical read-copy update in the linux kernel,'' in \emph{2018
  Design, Automation \& Test in Europe Conference \& Exhibition (DATE)}.\hskip
  1em plus 0.5em minus 0.4em\relax IEEE, 2018, pp. 61--66.

\bibitem{tassarotti2015verifying}
J.~Tassarotti, D.~Dreyer, and V.~Vafeiadis, ``Verifying read-copy-update in a
  logic for weak memory,'' \emph{ACM SIGPLAN Notices}, vol.~50, no.~6, pp.
  110--120, 2015.

\bibitem{dice2019bravo}
D.~Dice and A.~Kogan, ``Bravo---biased locking for reader-writer locks,'' in
  \emph{2019 $\{$USENIX$\}$ Annual Technical Conference
  ($\{$USENIX$\}$$\{$ATC$\}$ 19)}, 2019, pp. 315--328.

\bibitem{passiverwl}
\BIBentryALTinterwordspacing
R.~Liu, H.~Zhang, and H.~Chen, ``Scalable read-mostly synchronization using
  passive {Reader-Writer} locks,'' in \emph{2014 USENIX Annual Technical
  Conference (USENIX ATC 14)}.\hskip 1em plus 0.5em minus 0.4em\relax
  Philadelphia, PA: USENIX Association, Jun. 2014, pp. 219--230. [Online].
  Available:
  \url{https://www.usenix.org/conference/atc14/technical-sessions/presentation/liu}
\BIBentrySTDinterwordspacing

\bibitem{nir2013numarwl}
I.~Calciu, D.~Dice, Y.~Lev, V.~Luchangco, V.~J. Marathe, and N.~Shavit,
  ``Numa-aware reader-writer locks,'' in \emph{Proceedings of the 18th ACM
  SIGPLAN symposium on Principles and practice of parallel programming}, 2013,
  pp. 157--166.

\bibitem{chronos}
\BIBentryALTinterwordspacing
R.~Kapoor, G.~Porter, M.~Tewari, G.~M. Voelker, and A.~Vahdat, ``Chronos:
  Predictable low latency for data center applications,'' in \emph{Proceedings
  of the Third ACM Symposium on Cloud Computing}, ser. SoCC '12.\hskip 1em plus
  0.5em minus 0.4em\relax New York, NY, USA: Association for Computing
  Machinery, 2012. [Online]. Available:
  \url{https://doi.org/10.1145/2391229.2391238}
\BIBentrySTDinterwordspacing

\bibitem{shao-sdn-nsdi}
\BIBentryALTinterwordspacing
H.~Shao, X.~Wang, Y.~Lu, Y.~Yu, S.~Zheng, and Y.~Zhao, ``Accessing cloud with
  disaggregated {Software-Defined} router,'' in \emph{18th USENIX Symposium on
  Networked Systems Design and Implementation (NSDI 21)}.\hskip 1em plus 0.5em
  minus 0.4em\relax USENIX Association, Apr. 2021, pp. 1--14. [Online].
  Available: \url{https://www.usenix.org/conference/nsdi21/presentation/shao}
\BIBentrySTDinterwordspacing

\bibitem{AjitMvrlu}
\BIBentryALTinterwordspacing
J.~Kim, A.~Mathew, S.~Kashyap, M.~K. Ramanathan, and C.~Min, ``Mv-rlu: Scaling
  read-log-update with multi-versioning,'' in \emph{Proceedings of the
  Twenty-Fourth International Conference on Architectural Support for
  Programming Languages and Operating Systems}, ser. ASPLOS '19.\hskip 1em plus
  0.5em minus 0.4em\relax New York, NY, USA: Association for Computing
  Machinery, 2019, pp. 779--792. [Online]. Available:
  \url{https://doi.org/10.1145/3297858.3304040}
\BIBentrySTDinterwordspacing

\bibitem{GuoCrossley2017}
H.~Guo and P.~Crossley, ``Design of a time synchronization system based on gps
  and ieee 1588 for transmission substations,'' \emph{IEEE Transactions on
  Power Delivery}, vol.~32, no.~4, pp. 2091--2100, 2017.

\bibitem{Cosmos}
\BIBentryALTinterwordspacing
D.~Raychaudhuri, I.~Seskar, G.~Zussman, T.~Korakis, D.~Kilper, T.~Chen,
  J.~Kolodziejski, M.~Sherman, Z.~Kostic, X.~Gu, H.~Krishnaswamy,
  S.~Maheshwari, P.~Skrimponis, and C.~Gutterman, \emph{Challenge: COSMOS: A
  City-Scale Programmable Testbed for Experimentation with Advanced
  Wireless}.\hskip 1em plus 0.5em minus 0.4em\relax New York, NY, USA:
  Association for Computing Machinery, 2020. [Online]. Available:
  \url{https://doi.org/10.1145/3372224.3380891}
\BIBentrySTDinterwordspacing

\bibitem{Gallenmuller-mindthegap}
P.~Emmerich, S.~Gallenmüller, G.~Antichi, A.~W. Moore, and G.~Carle, ``Mind
  the gap - a comparison of software packet generators,'' in \emph{2017
  ACM/IEEE Symposium on Architectures for Networking and Communications Systems
  (ANCS)}, 2017, pp. 191--203.

\end{thebibliography}

\end{document}